\documentclass[aps,prb,twocolumn,groupedaddress,showpacs,amsmath,amssymb]{revtex4}

\usepackage{graphicx}
\usepackage{dcolumn}
\usepackage{bm}

\begin{document}

\title{Anomalous metallic state above the upper critical field of
the conventional three-dimensional superconductor AgSnSe$_{2}$ with
strong intrinsic disorder}

\date{\today}
\author{Zhi Ren}
\author{M. Kriener}
\author{A. A. Taskin}
\author{Satoshi Sasaki}
\author{Kouji Segawa}
\author{Yoichi Ando}
\email{y_ando@sanken.osaka-u.ac.jp}
\affiliation{Institute of Scientific and Industrial Research, Osaka University,
Ibaraki, Osaka 567-0047, Japan}

\begin{abstract}

We report superconducting properties of AgSnSe$_{2}$ which is a
conventional type-II superconductor in the very dirty limit due to
intrinsically strong electron scatterings. While this material is an
isotropic three-dimensional (3D) superconductor with a not-so-short
coherence length where strong vortex fluctuations are {\it not}
expected, we found that the magnetic-field-induced resistive transition
at fixed temperatures becomes increasingly broader toward zero
temperature and, surprisingly, that this broadened transition is taking
place largely {\it above} the upper critical field determined
thermodynamically from the specific heat. This result points to the
existence of an anomalous metallic state possibly caused by quantum
phase fluctuations in a strongly-disordered 3D superconductor.

\end{abstract}



\maketitle

\section{Introduction}

The interplay between disorder and superconductivity has attracted
sustained interest over the past few decades. In BCS superconductors,
according to Anderson's theorem, \cite{andersontheorem} the
superconducting critical temperature $T_{\rm c}$ and the energy gap
$\Delta$ remain unaffected by the presence of weak disorder, although in
unconventional superconductors both $T_{\rm c}$ and $\Delta$ are
strongly suppressed with weak disorder. \cite{Balian} In the strongly
disordered regime, things become interesting even in BCS
superconductors: In two-dimensional (2D) systems, strong disorder leads
to a universal superconductor-to-insulator transition at a critical
sheet resistance of $\sim h/4e^2$; \cite{Fisher, Haviland, Hebard,
Paalanen, Tanda, Fisher-1, Pang, Fisher-2} furthermore, it has been
found that, as $T$ tends to zero, there emerges a broad range of
magnetic field where the resistivity remains finite but is much smaller
than the normal-state value, which arguably signifies the importance of
quantum phase fluctuations. \cite{Mason,Steiner} In three-dimensional
(3D) systems, strong disorder leads to a disruption of the global
superconducting coherence through reductions in both the superfluid
density $n_{\rm s}$ and pairing interactions, which results in the
superconducting order parameter to spatially fluctuate and gives rise to
anomalous electronic states near the superconductor-to-metal transition.
\cite{Spivak,Feigel'man} Experimentally, while the 2D systems have been
actively studied in the past, \cite{Fisher, Haviland, Hebard, Paalanen,
Tanda, Fisher-1, Pang, Fisher-2, Mason, Steiner, Valles, Liu, Doniach,
Sambandamurthy, Dubi, Stewart, Sacepe} the role of strong disorder in
3D superconductors is just beginning to be addressed with a modern
viewpoint, \cite{Mondal,Chand} which naturally calls for explorations of
suitable materials for such a study.

In this paper, we report that AgSnSe$_2$, which is a low-$T_c$
conventional superconductor having a cubic structure, offers an ideal
playground for studying the effects of strong disorder in 3D
superconductors. We show that one can synthesize high-quality samples of
AgSnSe$_2$ presenting a very sharp superconducting transition, which
indicates that morphologically the system is homogeneous; nonetheless,
strong electron scatterings that are intrinsic to this system lead to an
extremely type-II superconductivity with the Ginzburg-Landau parameter
$\kappa_{\rm GL}$ of as large as 55. This causes the emergence of a
broad range of magnetic fields where the resistivity remains finite but
is much smaller than the normal-state value as $T \rightarrow 0$, which
is very much reminiscent of the behavior observed in disordered 2D
superconductors \cite{Mason,Steiner} or in quasi-2D systems like
high-$T_c$ cuprates. \cite{Osofsky,Mackenzie} Furthermore, such an
anomalous metallic state is found in the magnetic field range {\it
above} the upper critical field determined thermodynamically from the
specific heat. Since this anomalous state can be easily reached with the
magnetic field of 3 T and the microscopic mechanism of superconductivity
is well understood in conventional superconductors, future studies of
AgSnSe$_2$ would help elucidate the roles of quantum phase fluctuations
which remain controversial to date. \cite{Mason, Steiner, Spivak,
Feigel'man, Valles, Liu, Doniach, Sambandamurthy, Dubi, Stewart, Sacepe,
Mondal, Chand}

\section{$\mathbf{AgSnSe_2}$ Superconductor}

The superconductivity in AgSnSe$_{2}$ was discovered by Johnston and
Adrian \cite{ASS} in 1977 as a possible valence-skipping superconductor.
In valence-skipping compounds, the nominal valence state of a
constituent element happens to be the skipped valence for that element.
This characteristic has been of particular interest for
superconductivity, since it may provide a means to realize the so-called
negative-$U$ mechanism to enhance the superconducting transition
temperature $T_c$. \cite{varma} Well-known examples of such
valence-skipping superconductors are BaBi$_{1-x}$Pb$_{x}$O$_{3}$,
\cite{BPBO} Ba$_{1-x}$K$_{x}$BiO$_{3}$, \cite{BKBO} and
Pb$_{1-x}$Tl$_{x}$Te. \cite{Tl-PbTe} In all these compounds, $T_c$ is
relatively high for their low carrier density. At the same time, valence
fluctuations lead to significant electron scattering, making those
superconductors highly disordered even though the matrix is homogeneous.

AgSnSe$_{2}$ adopts the face-centered cubic structure, in which the
cation sites are randomly occupied by equal amount of Ag and Sn atoms.
Since the nominal valence of Sn is 3+ which coincides with its skipped
valence state, the Sn ions should be separated into 1:1 mixture of
Sn$^{2+}$ and Sn$^{4+}$ and the chemical formula of this material can be
more appropriately expressed as
(Ag$^{1+}$)(Sn$^{2+}$)$_{0.5}$(Sn$^{4+}$)$_{0.5}$(Se$^{2-}$)$_{2}$, as
evidenced by the measurements of the magnetic susceptibility \cite{ASS}
and the $^{119}$Sn M\"{o}ssbauer spectra. \cite{ASSmossbauer} Despite
the valence skipping nature and the carrier density of the order of
10$^{22}$ cm$^{-3}$, its $T_c$ of 4.7 K is not particularly high, which
suggests that the negative-$U$ mechanism is not at work here. Therefore,
in this compound one can explore manifestations of strong disorder
associated with valence fluctuations in the context of conventional
superconductivity without being bothered by strong electron
interactions.

\begin{figure}
\includegraphics*[width=8.7cm]{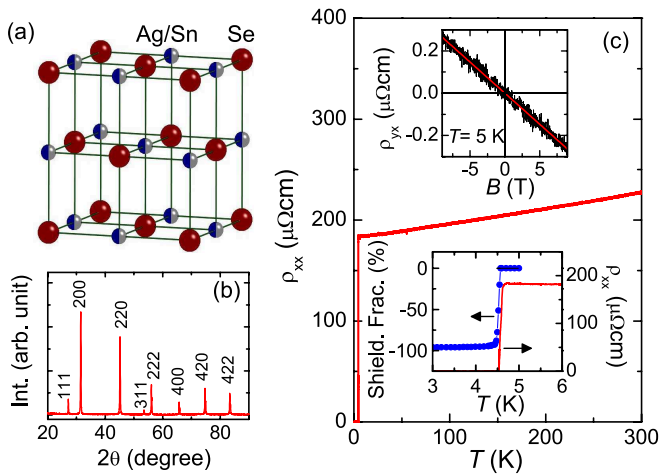}
\caption{(Color online)
(a) Face-centered cubic structure of AgSnSe$_{2}$. Note that Ag
and Sn atoms are randomly distributed in the cation sites. (b) XRD 
pattern of the AgSnSe$_{2}$ sample. All the
diffraction peaks can be well indexed using a cubic cell with the
F\emph{m}3\emph{m} space group. (c) Temperature dependence of
$\rho_{xx}$ of the AgSnSe$_{2}$ sample. Lower inset shows an enlarged
view of the data near the superconducting transition for $\rho_{xx}$
(right axis) and for the dc magnetic susceptibility measured upon
zero-field-cooling (left axis, shown in terms of the shielding fraction). 
Upper inset shows the magnetic-field dependence of
$\rho_{yx}$ at 5 K; its slope gives $n_e$ = 2.0 $\times$ 10$^{22}$ cm$^{-3}$.
}
\label{fig1}
\end{figure}

\section{Experimental}

We synthesized high-quality ingots of polycrystalline AgSnSe$_{2}$ by
using a two-step method as described in Ref. \onlinecite{ASS}. First,
high-purity shots of Ag (99.999\%), Sn (99.99\%), and Se (99.999\%) with
the stoichiometric ratio of 1:1:2 were melted in sealed evacuated quartz
tube at 800 $^{\circ}$C for 48 h with intermittent shaking to ensure
homogeneity, followed by quenching to room temperature. The quartz tube
containing the melt-quenched ingot was subsequently annealed at 450
$^{\circ}$C for two weeks, and then quenched into cold water. The
structure of the resulting sample was characterized by powder X-ray
diffraction (XRD). As can be seen in Fig. 1(b), the XRD pattern shows
sharp diffraction peaks consistent with the face-centered cubic
structure, giving the lattice parameter $a$ of 5.675(1) {\AA} which
agrees with the literature. \cite{ASS}

The ingot was cut into bar-shaped samples, and the temperature-dependent
magnetization $M$ under various magnetic fields was measured with a
vibrating sample magnetometer (VSM), while the magnetization curves at
fixed temperatures were measured with a commercial SQUID magnetometer
(Quantum Design MPMS- 1). The demagnetization effect was corrected for
by considering the slab geometry. \cite{demag} The resistivity
$\rho_{xx}$ and the Hall resistivity $\rho_{yx}$ were measured by using
a standard six-probe method where the contacts were made by spot-welding
gold wires. The specific heat $c_p$ was measured with a
relaxation-time method using Quantum Design PPMS-9. For consistency
reasons, all the data presented here were measured on exactly
the same sample.

\section{Basic characterizations}

\subsection{Transport properties and superconducting transition}

Figure 1(c) shows the temperature dependence of $\rho_{xx}$, which
exhibits only a weak metallic temperature dependence. This is
qualitatively similar to that reported \cite{ASS} for
Ag$_{0.76}$Sn$_{1.24}$Se$_{2}$ and signifies the presence of
intrinsically strong disorder that does not change much with the Ag/Sn
ratio. As shown in the upper inset of Fig. 1(c), $\rho_{yx}$ is linear
in $B$ and its slope at 5 K corresponds to an electron density $n_{\rm
e}$ of 2.0 $\times$ 10$^{22}$ cm$^{-3}$. The superconducting transition
in $\rho_{xx}$ occurs sharply between 4.5 and 4.6 K, and the magnetic
susceptibility shows the onset $T_c$ of 4.55 K, which corresponds to the
midpoint of the resistive transition. The superconducting shielding
faction achieves nearly 100\% after the demagnetization correction.
There is no resistivity upturn at low temperatures, giving no clear sign
of Anderson localization.

\begin{figure}
\includegraphics*[width=8.7cm]{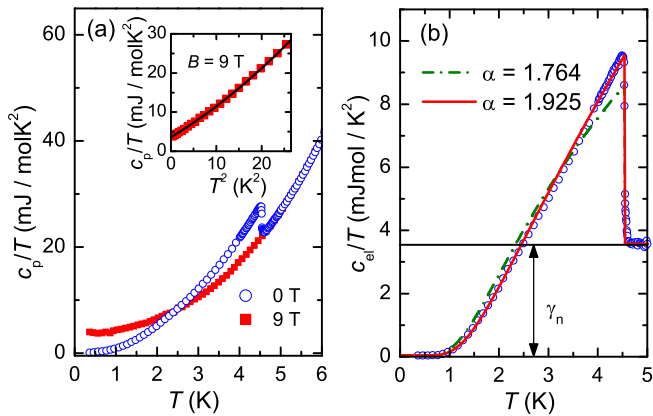}
\caption{(Color online)
(a) Temperature dependences of $c_p/T$ measured in 0 and 9 T.
Inset shows $c_p/T$ plotted as a function of $T^{2}$ for the 9-T
data, where the superconductivity is completely suppressed; the solid
line presents the best fit to the data using the Debye model (see Ref.
\onlinecite{NormalState}).
(b) Temperature dependence of $c_{\rm el}/T$ in 0 T. The dash-dotted line
is the calculated $c_{\rm el}/T$ curve given by the weak-coupling BCS
theory; the solid line is the best fit using a modified BCS
theory with $\alpha$ = 1.925. The horizontal solid line denotes
$\gamma_{\rm n}$.
}
\label{fig2}
\end{figure}

\subsection{Specific heat}

Figure 2(a) shows the temperature dependences of $c_p/T$ measured under
0 and 9 T. Application of a 9 T field completely suppresses the
superconductivity, enabling us to determine the phononic contribution
and the normal-state parameters such as the electronic specific-heat
coefficient $\gamma_{\rm n}$ and the effective mass $m^{\ast}$.
\cite{NormalState} Figure 2(b) shows the temperature dependence of the
electronic specific heat $c_{\rm el}/T$ in 0 T obtained by subtracting
the phononic contribution determined from the 9-T data. One can see that
$c_{\rm el}/T$ quickly approaches zero at low $T$, indicating a
fully-gapped nature. As for the precise temperature dependence, although
the simple weak-coupling BCS model (dash-dotted line) fails to describe
the $c_{\rm el}/T$ data, a modified BCS model \cite{alphamodel} that
allows the coupling constant $\alpha \equiv \Delta(0)/k_{\rm B}T_{\rm
c}$ to vary [$\Delta(0)$ is the gap at 0 K] can fit our data well with
$\alpha$ = 1.925 (solid line), which is only slightly larger than the
weak-coupling BCS value of 1.764. 

Using $m^{\ast}$ and $\alpha$ obtained from $c_{\rm p}(T)$,
\cite{NormalState} the Pippard coherence length $\xi_{\rm 0}$ can be
estimated as $\xi_{\rm 0}$ = $\hbar^{2}(3\pi^{2}n_{e})^{1/3}/(\pi\alpha
k_{\rm B}T_{\rm c}m^{\ast})$ = 181 nm. This is to be contrasted with the
mean free path $\ell$ = $\hbar(3\pi^{2})^{1/3}/(\rho_{\rm
0}n_{e}^{2/3}e^{2})$ = 0.9 nm, where $\rho_{\rm 0}$ is the residual
resistivity. This $\ell$ is less than $2a$ and is very short, pointing
to the existence of strong scattering which is consistent with the
valence fluctuations of Sn. The ratio $l$/$\xi_{\rm 0}$ = 0.005 means
that AgSnSe$_{2}$ is in the very dirty limit. Yet, $k_{\rm F}\ell$ is
estimated to be $\sim$8 and hence the system is still a good metal. Note
that the estimations of $m^{\ast}$, $k_{\rm F}$, $\xi_{\rm 0}$, and
$\ell$ in this paper are based on the free electron theory; once the
band structure of AgSnSe$_{2}$ is known, those estimations can be made
more accurate.

\begin{figure}
\includegraphics*[width=8.7cm]{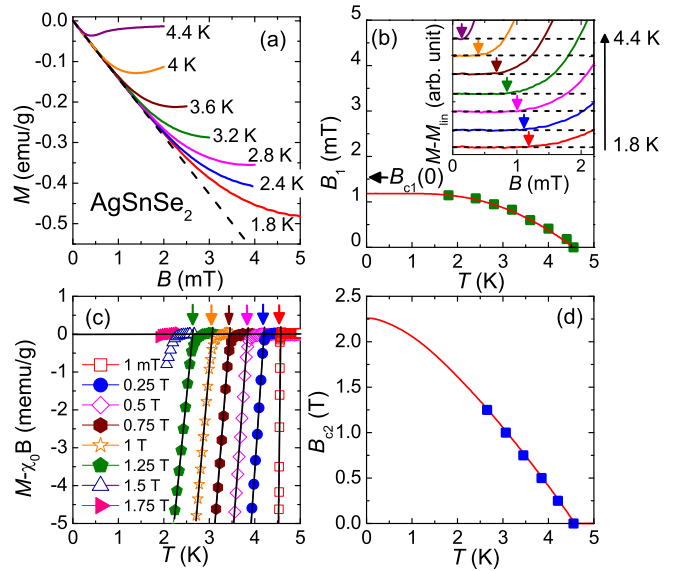}
\caption{(Color online)
(a) $M(B)$ curves measured after zero-field cooling to various
temperatures. The dashed line marks the initial linear behavior. (b)
Plot of $B_{\rm 1}$ vs $T$; the solid line is a fit to the
data using the local dirty limit formula. The $B_{\rm c1}$(0)
obtained after the demagnetization correction is marked by an
arrow. Inset shows the $B$ dependences of $\Delta M$ $\equiv$ $M - M_{\rm
lin}$, where $M_{\rm lin}$ denotes the initial Meissner contribution;
the data are shifted vertically for clarity. The determined
$B_{1}$ value for each temperature is shown by arrows. (c) Temperature
dependences of $M_{\rm sc} \equiv M - \chi_{0} B$ under various magnetic
fields. The $T_{\rm c}^{\rm Mag}(B)$ values are indicated by arrows. (d) $B_{\rm
c2}^{\rm Mag}$ vs $T$ phase diagram; the solid line is a WHH fit to the data.
}
\label{fig3}
\end{figure}

\subsection{Critical fields}

Figure 3 summarizes the results of the magnetization measurements to
extract the values of the upper and lower critical fields. Figure 3(a)
shows $M(B)$ curves measured after zero-field cooling to various
temperatures. We define $B_{\rm 1}$ at each temperature as the value at
which the $M(B)$ data deviates from its initial linear behavior, as can
be more clearly seen in the inset of Fig. 3(b). Extrapolation \cite{B1}
of the resulting $B_{\rm 1}(T)$ data to $T$ = 0 K yields $B_{\rm 1}(0)$
= 1.18 mT which, after correcting for the demagnetization effect,
\cite{demag} gives the lower critical field at $T$ = 0 K, $B_{\rm
c1}(0)$, of 1.58 mT.

Figure 3(c) shows the temperature dependences of the magnetization near
the superconducting transition under various magnetic fields. A
paramagnetic background with a temperature-independent susceptibility
$\chi_{0}$ = 4.2 $\times$ 10$^{-8}$ emu/g has been subtracted from the
data, \cite{chi0note} and what is plotted is $M_{\rm sc} \equiv M -
\chi_{0}B$. A gradual suppression of superconductivity with increasing
magnetic field is evident. For each magnetic field, we determined the
onset temperature of the transition, $T_{\rm c}^{\rm Mag}(B)$, from the
intersection of the linear extrapolation of the $M_{\rm sc}(T)$ data to
the $M_{\rm sc}$ = 0 line, as indicated by the arrows in Fig. 3(c). The
obtained $T_{\rm c}^{\rm Mag}(B)$ data shown in Fig. 3(d) give the
temperature dependence of the magnetically-determined upper critical
field, $B_{c2}^{\rm Mag}(T)$, which is extrapolated to $T$ = 0 K using the
Werthamer-Helfand-Hohenberg (WHH) theory \cite{WHH} to yield $B_{\rm
c2}^{\rm Mag}(0)$ = 2.26 T.

\begin{figure}
\includegraphics*[width=6cm]{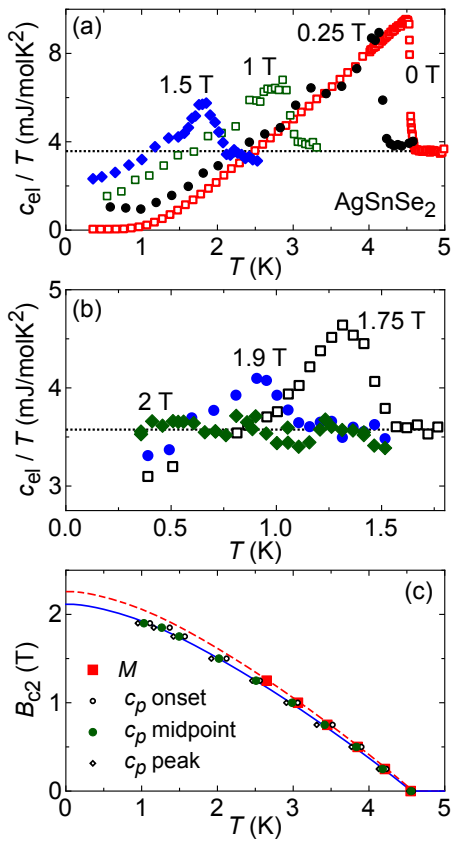}
\caption{(Color online)
(a, b) Superconducting transition measured in the temperature dependence 
of the specific heat under various magnetic fields.
(c) $B_{c2}^{Cp}$ vs $T$ phase diagram; the solid line is a WHH fit to the data.
For comparison, $B_{c2}^{\rm Mag}(T)$ and its WHH fitting are shown with filled
squares and a dashed line, respectively.
}
\label{fig4}
\end{figure}

To evaluate the upper critical field at low temperatures in a more
direct way, we measured the superconducting transition in specific heat
under various magnetic fields. Figure 4(a) shows representative data of
$c_{\rm el}/T$ vs $T$ measured in 0, 0.25, 1.0, and 1.5 T, where the
superconducting transition is well resolved. In higher magnetic fields,
the specific-heat anomaly becomes weaker, as shown in Fig. 4(b) for
1.75, 1.9, and 2.0 T; here, the transition can still be seen in 1.9 T,
but no clear anomaly is discernible at 2.0 T. These data demonstrate
that the upper critical field is thermodynamically well-defined in
AgSnSe$_{2}$ at low temperature and that it lies around 2 T near $T$ = 0
K.

To quantitatively analyze the superconducting transition in the
specific-heat data, for each magnetic field we determined three
characteristic temperatures: onset of the transition (below which the
specific heat grows), midpoint of the transition, and the peak in
$c_p(T)$. It is useful to note that the transition width $\Delta
T_c^{Cp}$, which we define by the difference between the onset and peak
temperatures, remains reasonably narrow with $\Delta T_c^{Cp} \lesssim
0.3$ K at all temperatures. We choose to use the midpoint of the
specific-heat transition for defining the transition temperature $T_{\rm
c}^{Cp}(B)$ for each magnetic field, which in turn gives the temperature
dependence of the upper critical field $B_{c2}^{Cp}(T)$. 

The solid line in Fig. 4(c) is the fitting of the WHH theory to the
$B_{c2}^{Cp}(T)$ data, which gives the $T$ = 0 K value $B_{c2}^{Cp}(0)$
of 2.11 T. For comparison, we reproduced in Fig. 4(c) the $B_{c2}^{\rm
Mag}(T)$ data obtained from the magnetization measurements. The
difference in the WHH fittings for the specific-heat and magnetization
data is only 7\%, which indicates that the two thermodynamical
measurements are essentially consistent. In the following discussions,
we take $B_{c2}^{Cp}$ as the thermodynamically-determined $B_{c2}$.

\subsection{Superconducting parameters}

Now we estimate various superconducting parameters of AgSnSe$_{2}$ by
using $B_{\rm c2}^{Cp}(0)$ as the zero-temperature upper critical
field $B_{\rm c2}(0)$. Note that since AgSnSe$_{2}$ has cubic symmetry,
no anisotropy is expected. The Ginzburg-Landau (GL) coherence length
$\xi_{\rm GL}(0)$ is estimated as $\xi_{\rm GL}$(0) =
$\sqrt{\Phi_{0}/2\pi B_{\rm c2}(0)}$ = 12.5 nm, where $\Phi_{0}$ = 2.07
$\times$ 10$^{-15}$ Wb is the flux quantum. Alternatively, one
may also estimate $\xi_{\rm GL}$ from $\xi_{\rm GL}(0)$ $\simeq$
$\sqrt{\xi_{\rm 0}\ell}$ for dirty superconductors, which yields
$\xi_{\rm GL}$ $\simeq$ 12.7 nm. The consistency is gratifying. The
equation \cite{Hu} $B_{\rm c1}$(0)/$B_{\rm c2}$(0) = $(\ln \kappa_{\rm
GL} + 0.5)/(2\kappa_{\rm GL}^{2})$ gives the GL parameter $\kappa_{\rm
GL}$ = 55, which in turn allows us to calculate the effective
penetration depth $\lambda_{\rm eff}$ = 685 nm through $B_{\rm c1}(0)$ =
$(\Phi_{0}/4\pi\lambda_{\rm eff}^{2})(\ln \kappa_{\rm GL}+0.5)$. This gives
the London penetration depth $\lambda_{\rm L}$ $\simeq$ $\lambda_{\rm
eff}\sqrt{\ell/\xi_{\rm 0}}$ = 48.3 nm.

Since AgSnSe$_{2}$ is a very dirty superconductor, the superfluid
density $n_{s}$ is expected to be reduced by a factor of $\sim$
$\ell/\xi_{\rm 0}$. \cite{Tinkham} Indeed, the calculation of $n_{s}$
based on $n_{s}$ = $m^{\ast}/(\mu_{0}e^{2}\lambda_{\rm eff}^{2})$ yields
$n_{s}$ $\approx$ 9 $\times$ 10$^{19}$ cm$^{-3}$, which is close to the
estimation of $n_{s}$ $\approx$ $n_{\rm e}(\ell/\xi_{\rm 0})$ = 1
$\times$ 10$^{20}$ cm$^{-3}$, reassuring the applicability of the BCS
theory to AgSnSe$_{2}$ and the soundness of our $B_{c1}$ measurements
which can be adversely affected by flux pinning.

\begin{figure}
\includegraphics*[width=8cm]{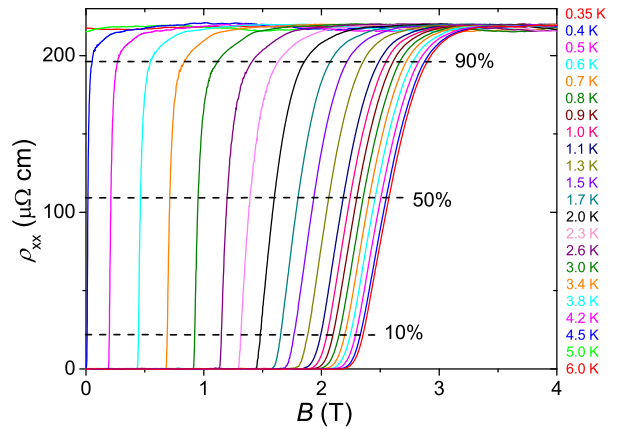}
\caption{(Color online)
Magnetic-field-induced resistive transition measured at various
temperatures. The level of $\rho_{xx}$ that is used for determining
$B_{10\%}$, $B_{50\%}$, and $B_{90\%}$ are indicated by
dashed lines.
}
\label{fig5}
\end{figure}

\section{Anomalous Metallic State}

\subsection{Resistive transition in magnetic fields}

Despite the conventional nature of the superconductivity in
AgSnSe$_{2}$, nontrivial physics becomes evident when one looks at the
magnetic-field-induced resistive transition at fixed temperatures (Fig.
5). One can easily see that the resistive transition is very sharp at
temperatures close to $T_c$, but it becomes increasingly broader at
lower temperatures. We determined the values of the magnetic field at
which $\rho_{xx}$ recovers a certain percentage of the normal-state
resistivity $\rho_{\rm N}$ for each temperature, and denote them
$B_{10\%}(T)$, $B_{50\%}(T)$, {\it etc.} We also determined the
irreversibility field $B_{\rm irr}(T)$ to mark the onset of a finite
resistivity within our sensitivity limit, which corresponds to 0.15\% of
$\rho_{\rm N}$. The results are shown in Fig. 6, where we also plotted
the thermodynamically-determined $B_{c2}(T)$ with a solid line.

Usually, either $B_{50\%}(T)$ or $B_{90\%}(T)$ are considered to give a
measure of the upper critical field. \cite{Ando,Tenhover} In the present
case, for temperatures down to $T/T_c \sim 1/2$, $B_{50\%}(T)$ agrees
reasonably well with the thermodynamically determined $B_{c2}$, but at
lower temperatures the resistive transition occurs largely above
$B_{c2}$. This means that at low temperature, there is a wide range of
magnetic field above $B_{c2}$ where the resistivity is finite but is
noticeably smaller than $\rho_{\rm N}$. The existence of such an
anomalous metallic state above $B_{c2}$ has never been documented for a
conventional 3D superconductor, and this is the main finding of the
present work. Note that the Pauli paramagnetic limit is estimated to be
8.4 T, which is well above the anomalous metallic regime observed here. 

\begin{figure}
\includegraphics*[width=6.5cm]{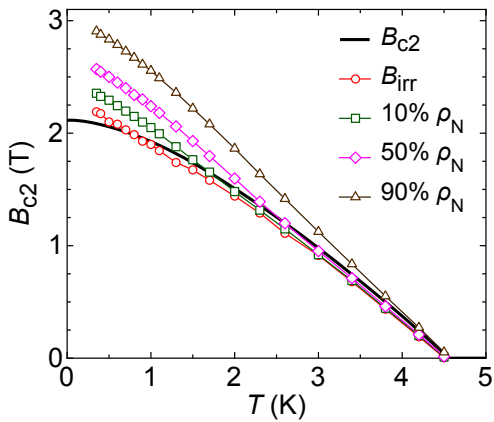}
\caption{(Color online)
Plots of $B_{c2}$ determined thermodynamically from specific-heat 
measurements (solid
line) together with $B_{\rm irr}$, $B_{10\%}$, $B_{50\%}$, and
$B_{90\%}$ obtained from the resistive transition data. 
}
\label{fig6}
\end{figure}

Similar broadening of the magnetic-field-induced resistive transition at
low temperatures has been observed in high-temperature superconductors
such as Bi$_2$Sr$_2$CuO$_y$, \cite{Osofsky} Tl$_2$Ba$_2$CuO$_y$,
\cite{Mackenzie} MgB$_2$, \cite{Szabo} and Ba$_{1-x}$K$_x$BiO$_3$.
\cite{Samuely} This phenomenon is generally believed to be caused by
enhanced vortex fluctuations and the resulting vortex liquid phase due
to a short coherence length that is inherent to high-temperature
superconductors. In light of this common belief, the present observation
in AgSnSe$_{2}$ is striking. Here the coherence length is not so short,
12.1 nm, and the system is 3D, both of which work against strong vortex
fluctuations. Therefore, our result calls for a reinterpretation of the
low-temperature broadening of the magnetic-field-induced resistive
transition.

\subsection{Quantum phase fluctuations}

To understand the above results, an interesting possibility is the
scenario proposed by Spivak, Oreto and Kivelson \cite{Spivak} that the
strong disorder leads to formations of superconducting ``puddles" that
are Josephson-coupled to each other, and quantum fluctuations of the
relative phase between those puddles give rise to a broad range of
magnetic fields where the resistivity remains finite and smaller than
$\rho_{\rm N}$ as $T \rightarrow 0$. If this is indeed the case, our
data give evidence for the important role of quantum phase fluctuations
in disordered 3D conventional superconductors. Since the
superconductivity mechanism is well understood here (as opposed to the
case of high-temperature superconductors) and the required magnetic
field to access the quantum fluctuation regime is low (only $\sim$3 T),
AgSnSe$_{2}$ provides a convenient platform for studying the role of
quantum fluctuations in the magnetic-field-induced
superconductor-to-metal transition.

Another anomalous behavior observed here is that at low temperatures,
not only $B_{90\%}$ but also $B_{\rm irr}$ show little tendency toward
saturation. A similar behavior was previously observed in
strongly-disordered amorphous superconductors \cite{Tenhover} and was
discussed to be due to the weak-localization effect. \cite{Coffey}
However, the discrepancy between the resistive transition and the
thermodynamically-determined $B_{c2}$ makes the conclusion of those old
studies \cite{Tenhover,Coffey} questionable. Hence, the physical meaning
of the resistive transition in disordered superconductors had better be
reconsidered by taking into account the important role of quantum phase
fluctuations.

\section{Summary}

We have fully characterized the AgSnSe$_{2}$ superconductor where its
valence-skipping nature leads to intrinsically strong electron
scattering without any obvious enhancement of $T_c$. At low temperature,
we observed an anomalous broadening of the magnetic-field-induced
resistive transition which occurs largely above the
thermodynamically-determined $B_{c2}$. This means the existence of an
anomalous metallic state above $B_{c2}$ where quantum phase
fluctuations associated with a magnetic-field-induced
superconductor-to-metal transition are likely to be playing an important
role. This makes AgSnSe$_{2}$ a useful platform for studying the role of
quantum fluctuations in 3D disordered superconductors.

\begin{acknowledgments}
We thank S. A. Kivelson for illuminating discussions.
This work was supported by JSPS (NEXT Program),
MEXT (Innovative Area ``Topological Quantum Phenomena" KAKENHI
22103004), and AFOSR (AOARD 124038).
\end{acknowledgments}


\begin{thebibliography}{99}
\bibitem{NormalState}
As can be seen in the inset of Fig. 2(a), the 9 T data below 5 K can be
well described by the Debye model $c_p/T$ = $\gamma_{\rm n}$ +
$A_{\rm 3}$$T^{2}$ + $A_{\rm 5}T^{4}$, where $\gamma_{\rm n}$ is the
normal-state electronic specific-heat coefficient and $A_{i}$ ($i$ = 3,
5) denote the coefficients of the phononic contribution. The obtained
parameters are $\gamma_{\rm n}$ = 3.5 mJ/molK$^{2}$, $A_{\rm 3}$ = 0.7
mJ/molK$^{4}$, and $A_{\rm 5}$ = 0.009 mJ/molK$^{6}$. The $A_{\rm 3}$ value
gives the Debye temperature $\Theta_{\rm D}$ = 220 K via
$\Theta_{\rm D}$ = (12$\pi$$^{4}$$nR$/5$A_{\rm 3}$)$^{1/3}$ with $n$ =
4 and $R$ = 8.314 J/molK. In addition, if one assumes
a single spherical Fermi surface, the effective mass
$m^{\ast}$ can be estimated as $m^{\ast}$ = $3\hbar^{2}\gamma_{\rm
n}/(V_{\rm mol}k_{\rm B}^{2}(3\pi^{2}n_{e})^{1/3})$ $\simeq$
1.5$m_{e}$, where $V_{\rm mol}$ = 55 cm$^{3}$/mol and $m_{e}$ is the
free electron mass.
\bibitem{B1}
We fit the $B_{\rm 1}(T)$ curve with
$B_{\rm 1}$ $\propto$ $1/\lambda_{\rm eff}^{2}$ $\propto$
$[\Delta(T)/\Delta(0)]\tanh(\Delta(T)/2k_{\rm B}T)$ for the local dirty
limit, \cite{Tinkham} where $\lambda_{\rm eff}$ is the effective
penetration depth.
\bibitem{chi0note}
The positive $\chi_{0}$ is probably due to the presence of a very small
amount of paramagnetic impurities.
\end{thebibliography}
\end{document}